\documentclass[aps,prl,twocolumn,floatfix,longbibliography,a4paper,superscriptaddress]{revtex4-1}

\usepackage{amsmath, amssymb}
\usepackage{mathtools}
\usepackage{lipsum}
\usepackage{times}
\usepackage{graphicx}
\usepackage{wasysym}
\usepackage{bm}
\usepackage{hyperref}
\usepackage{xspace}
\usepackage{siunitx}
\usepackage{color}
\usepackage[labelsep=space,labelfont={color=myColor,bf}]{caption}
\usepackage{braket}
\usepackage{todonotes}
\presetkeys{todonotes}{size=\footnotesize, color=yellow!90}{}

\definecolor{myColor}{rgb}{0.02,0.12,0.3}
\definecolor{myciteColor}{rgb}{0.39,0.7,0.89}
\hypersetup{colorlinks=true, linkcolor=myColor, filecolor=myColor, urlcolor=myColor, citecolor=myColor, urlcolor=myColor}
\graphicspath{{./Figures/}}

\DeclareSIUnit{\nK}{\nano\kelvin}
\DeclareSIUnit{\aB}{\emph{a}_0}
\DeclareSIUnit{\G}{G}

\renewcommand{\figurename}[1]{Fig.~}

\newcommand{\kB}{k_{\text{B}}}

\newcommand{\cB}{c_{\text{B}}}
\newcommand{\oB}{\omega_{\text{B}}}

\newcommand{\g}{\tilde{g}}

\newcommand{\ns}{n_{\text{s}}}

\newcommand{\Dc}{\mathcal{D}_{\text{c}}}
\newcommand{\Tc}{T_{\text{c}}}

\newcommand{\Ds}{\mathcal{D}_{\text{s}}}
\newcommand{\oz}{\omega_{z}}

\newcommand{\Lx}{L_{x}}
\newcommand{\Ly}{L_{y}}
\newcommand{\kT}{\kappa_{T}}
\newcommand{\kS}{\kappa_{S}}

\newcommand{\G}{\Gamma_{\text{coll}}}
\newcommand{\xn}{\chi_{\text{nn}}}
\newcommand{\Ima}{{\text{Im}}}
\newcommand{\fsum}{f_{\text{sum}}}
\newcommand{\tA}{\tilde{A}}

\begin{document}


\title{Observation of first and second sound in a Berezinskii--Kosterlitz--Thouless superfluid}
\author{Panagiotis Christodoulou}
\author{Maciej Ga{\l}ka}
\author{Nishant Dogra}
\affiliation{Cavendish Laboratory, University of Cambridge, J. J. Thomson Avenue, Cambridge CB3 0HE, United Kingdom}
\author{Raphael Lopes}
\affiliation{Laboratoire Kastler Brossel, Collège de France, CNRS, ENS-PSL Research University, Sorbonne Universit\'e, 11 Place Marcelin Berthelot, 75005 Paris, France}
\author{Julian Schmitt}
\altaffiliation[Present address: ]{Institut f{\"u}r Angewandte Physik, Universit{\"a}t Bonn, Wegelerstra{\ss}e 8, 53115 Bonn, Germany}
\affiliation{Cavendish Laboratory, University of Cambridge, J. J. Thomson Avenue, Cambridge CB3 0HE, United Kingdom}
\author{Zoran Hadzibabic}
\affiliation{Cavendish Laboratory, University of Cambridge, J. J. Thomson Avenue, Cambridge CB3 0HE, United Kingdom}

\date{\today}

\maketitle

\newpage

\textbf{Superfluidity in its various forms has fascinated scientists since the observation of frictionless flow in liquid helium II~\citep{Kapitza:1938,Allen:1938}. In three spatial dimensions (3D), it is conceptually associated with the emergence of long-range order (LRO) at a critical temperature $\Tc$. One of its hallmarks, predicted by the highly successful two-fluid model~\citep{Tisza:1938,Landau:1941b} and 
observed in both liquid helium~\citep{Peshkov:1960} and ultracold atomic gases~\citep{Stamper-Kurn:1998a,Sidorenkov:2013,Zwierlein:private, Hilker:2020}, is the existence of two kinds of sound excitations, the first and second sound. In 2D systems, thermal fluctuations preclude LRO~\citep{Hohenberg:1967,Mermin:1966}, but superfluidity nevertheless emerges at a nonzero $\Tc$ via the infinite-order Berezinskii-Kosterlitz-Thouless (BKT) transition~\citep{Berezinskii:1971,Kosterlitz:1973}, which is associated with a universal jump in the superfluid density $\ns$~\citep{Nelson:1977} without any discontinuities in the fluid's thermodynamic properties. 
BKT superfluids are also predicted to support two sounds, but the observation of this has remained elusive. Here we observe first and second sound in a homogeneous 2D atomic Bose gas, and from the two temperature-dependent sound speeds extract its superfluid density~\citep{Prokofev:2001,Prokofev:2002,Ozawa:2014,Ota:2018b}. 
Our results agree with BKT theory, including the prediction for the universal superfluid-density jump.
}

The hydrodynamic two-fluid theory~\citep{Landau:1941b} models a fluid below $\Tc$ as a mixture of a superfluid component and a viscous normal component that carries all the entropy, and assumes that the two are in local thermodynamic equilibrium. 
The two sounds then correspond to different variations of the total density and the entropy per particle. In 3D, in the nearly-incompressible liquid helium the higher-speed first sound is a pure density wave and the lower-speed second sound is a pure entropy wave, but in general both sounds can involve both density and entropy variations~\citep{Hu:2010}. Since above $\Tc$ the normal fluid supports just the ordinary first-sound density wave, the appearance of the second sound mode is a striking manifestation of superfluidity.  

Despite the different nature of the phase transition, the two-sounds phenomenology is also expected for BKT superfluids. However, in liquid-helium films, where the BKT transition was first observed~\citep{Bishop:1978}, the (first and second) sound propagation is inhibited by the substrate roughness. On the other hand, in 2D atomic gases, 
where many complementary BKT experiments were performed~\citep{Hadzibabic:2006,Clade:2009,Tung:2010,Yefsah:2011,Hung:2011,Hadzibabic:2011,Desbuquois:2012,Ha:2013,Choi:2013,Chomaz:2015,Fletcher:2015, Murthy:2015,Ville:2018}, so far only one sound mode was seen.
In a weakly interacting Bose gas~\citep{Ville:2018} collisionless sound was observed~\citep{Ota:2018a,Cappellaro:2018} (see also~\citep{Wu:2020}) and showed no discontinuity at $\Tc$, while in a strongly interacting Fermi gas~\citep{Bohlen:2020} one pure density mode was observed well below $\Tc$.


\begin{figure}[t] 
  \centering
  \includegraphics[width=\columnwidth]{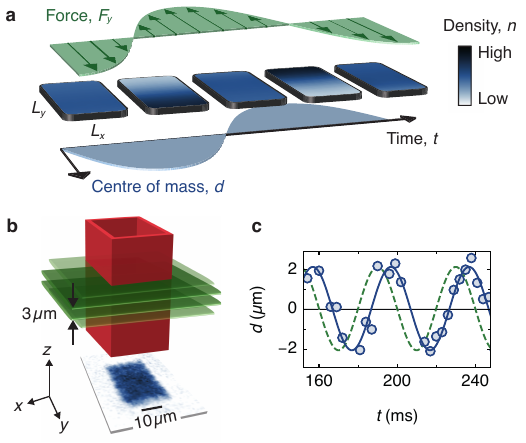}
 \caption{$\vert$ {\bf Sound excitations in a homogeneous 2D Bose gas.} {\bf a,} We apply an in-plane, spatially-uniform force $F_y(t) = F_0 \sin(\omega t)$, created by a magnetic field gradient, on a homogeneous optically trapped 2D gas. This excites long-wavelength density modulations with wavevector $q=\pi/\Ly$, which results in a displacement of the cloud's centre of mass, $d(t)$. On resonance, $d$ oscillates $\pi/2$ out of phase from $F_y$. {\bf b,} Outline of our trapping setup (see text), and an absorption image of our 2D gas. {\bf c,} An example of $d(t)$ oscillation, for a gas below $\Tc$ and $\omega/(2\pi) = 25$~Hz near the second-sound resonance; for comparison, $\Ly \approx 33~\mu{\rm m}$. The dashed curve indicates the phase of $F_y(t)$.}
\label{fig:1}
\end{figure} 

\begin{figure*}[t] 
   \centering
  \includegraphics[width=\textwidth]{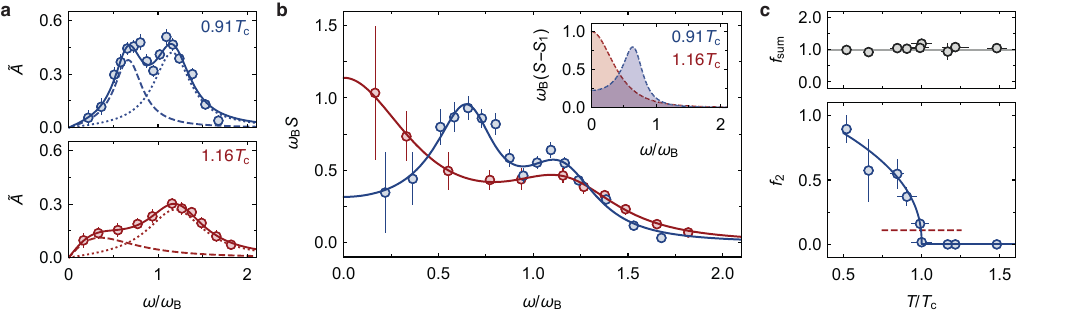}
 \caption{$\vert$ {\bf First and second sound.} 
 {\bf a,} Normalised response spectra $\tA(\omega)$ at two different temperatures, for $n \approx 3~\mu$m$^{-2}$, $\Ly \approx 33~\mu$m, and $F_0/m \approx 0.074~{\rm m}/{\rm s}^2$; $\oB$ is the Bogoliubov frequency (see text).
 Below $\Tc$ we observe two resonances corresponding to the first (dotted) and second (dashed) sound. Above $\Tc$ we instead observe just the first-sound resonance (dotted), while the second sound is replaced by a diffusive, overdamped mode (dashed).
{\bf b,} The corresponding dynamical structure factors $S(\omega)$. 
Here the diffusive mode peaks at $\omega =0$ and its distinction from the second-sound resonance is more visually striking. The inset shows the fitted contributions to $S(\omega)$ from the second sound below $\Tc$ (blue) and the diffusive mode above $\Tc$ (red), omitting for clarity the first-sound contributions $S_1(\omega)$, which are similar at the two temperatures.
{\bf c,} The $f$ sum rule and the critical point. 
Top panel: we verify the $f$ sum rule for a wide range of temperatures. Bottom: $f_2$, the second-sound contribution to the constant $\fsum$, vanishes with increasing $T$ and we use it to experimentally identify $\Tc$. The solid line is a heuristic power-law fit. The dashed line indicates the value of the theoretical discontinuity in $f_2$ at $\Tc$ in an infinite system (see text). 
 The error bars in all panels show standard fitting errors.   }
   \label{fig:2}
\end{figure*}

We observe both first and second sound in the long-wavelength density response of a homogeneous 2D Bose gas to an external driving force (see Fig.~\ref{fig:1}a). In our $^{39}$K gas of density $n \approx 3~\mu$m$^{-2}$ and characterised by a dimensionless interaction strength $\g = 0.64(3)$~\citep{Petrov:2000a,Hadzibabic:2011}, 
the elastic collision rate is sufficient for collisional hydrodynamic behaviour~\footnote{Near $\Tc$ the elastic collision rate is about four times larger than the first-sound frequency~\citep{Petrov:2000a,Ville:2018}} and the compressibility is still sufficiently high such that near $\Tc$ 
our driving force excites both sounds effectively~\citep{Ota:2018b,Hu:2010}.

Our homogeneous 2D gases are prepared in a node of a vertical 1D optical lattice (green in Fig.~\ref{fig:1}b) with harmonic trap frequency $\oz/(2 \pi) = 5.5(1)$~kHz; they are deep in the 2D regime, with both the interaction and thermal energy per particle below $0.3 \,\hbar \oz$, where $\hbar$ is the reduced Planck's constant.
In the $x-y$ plane, we confine the atoms to a rectangular box of size $\Lx \times \Ly$ and potential-energy wall height $U_0$, using a hollow laser beam (red in Fig.~\ref{fig:1}b); we tune $U_0/\kB$, where $\kB$ is the Boltzmann constant, between $100$ and $300$~nK to vary the gas temperature $T$
(see Methods).
We control the interaction strength $\g = \sqrt{8\pi m \oz/\hbar}\, a$, where $m$ is the atom mass and $a$ the scattering length~\citep{Hadzibabic:2011}, via a magnetic Feshbach resonance at $402.7$~G~\citep{Fletcher:2018}. Our $\g=0.64(3)$ corresponds to a relatively high $a = 522 (23)~a_0$, where $a_0$ is the Bohr radius, and comes with the cost of enhanced three-body losses, but $n$ stays within $15\%$ of its average value during the measurements. 

The driving force $F_y = F_0 \sin(\omega t)$ excites the longest-wavelength sound mode(s) in our box, with wavevector $q = \pi/\Ly$~\citep{Navon:2016}. This results in a density perturbation $\delta n(y,t)/n = b(t) \sin (\pi y/\Ly)$, with $y=0$ in the box centre, 
and displaces the cloud's centre of mass by $d(t) = 2b(t)\Ly/\pi^2$. We choose $F_0$ so that the maximal $d(t)$ is a few~\% of $\Ly$ (see Fig.~\ref{fig:1}c) and fit $d(t) = R(\omega) \sin(\omega t) - A(\omega) \cos(\omega t)$, which gives the reactive ($R$) and absorptive ($A$) response~\footnote{We find the same results whether we fit the density profiles to get $b(t)$ or simply sum over the images to get $d(t)$.}.
We focus on $A(\omega)$, which is proportional to the imaginary part of the density response function $\xn$ and satisfies the $f$ sum rule~\citep{Pitaevskii:2016}
\begin{equation}
    \fsum = \int_{-\infty}^{\infty} {\rm d}\omega \, \omega \frac{ \pi A(\omega)}{8 F_0/m} = 1  \,;
    \label{eq:f}
\end{equation}
moreover, it directly gives the dynamical structure factor $S(q,\omega) = \pi q^2 \kB T A(\omega)/(8\omega F_0)$~\citep{Hu:2010} (see Methods). 


Fig.~\ref{fig:2} shows two qualitatively different response spectra below and above the critical temperature, together with our experimental determination of $\Tc$ (Fig.~\ref{fig:2}c). Here we express our results in dimensionless form using the Bogoliubov frequency $\oB = \cB q$, where $\cB=\hbar\sqrt{n\g}/m \approx 2.3$~mm/s is the Bogoliubov sound speed. We define the dimensionless $\tA = \pi m \oB^2 A /(8F_0)$, so  $\fsum = \int {\rm d}\omega \,\omega  \tA/\oB^2$ and $S = \kB T \tA / (m\cB^2 \omega)$.
For our fitting procedure see Methods.

In Fig.~\ref{fig:2}a, below $\Tc$ we observe two resolved resonances corresponding to the first (dotted) and second (dashed) sound; the respective resonance frequencies $\omega_1$ and $\omega_2 < \omega_1$ give the two sound speeds $c_{1,2} = \omega_{1,2}/q$. 
On the other hand, above $\Tc$ we observe only the first-sound resonance (dotted). Here the low-frequency `shoulder' (dashed) is due to the  diffusive heat mode that replaces the second sound; this mode also couples to density fluctuations, and is in fact necessary to exhaust the $f$ sum rule.
Fig.~\ref{fig:2}b shows the corresponding dynamical structure factors $S(\omega)$, and the inset highlights the qualitative difference  between the propagating second sound ($c_2 >0$) and the diffusive mode, which peaks at $\omega=0$, corresponding to $c_2=0$~\citep{Hohenberg:1964}. 
From the diffusive-mode width we estimate the thermal diffusivity $D_{\text{T}} = 5(2) \hbar/m$. With a caveat that our sound resonances might be broadened by three-body atom loss, we note that their widths imply sound diffusivities $D_{\rm s, 1} = 7(1)\hbar/m$ and $10(2) \hbar/m$ for the first sound below and above $\Tc$, respectively, and $D_{\rm s, 2} = 6(1)\hbar/m$ for the second sound below $\Tc$. For comparison, several times lower sound diffusivities, $\sim \hbar/m$, were observed in strongly interacting 2D~\citep{Bohlen:2020} and 3D~\citep{Patel:2020} Fermi gases. 

\begin{figure*} [t] 
  \centering
  \includegraphics[width=\textwidth]{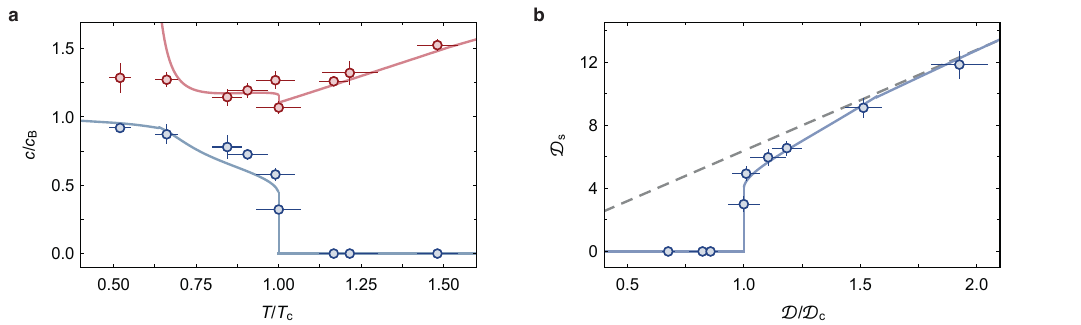}
 \caption{$\vert$ {\bf The sound speeds and the superfluid density.} 
{\bf a,} Normalised sound speeds, $c_1/\cB$ (red points) and $c_2/\cB$ (blue points), where $\cB$ is the Bogoliubov speed, and the corresponding theoretical predictions without any free parameters. The discontinuities in the theoretical speeds at $\Tc$ correspond to the infinite-system superfluid-density jump.
{\bf b,} The superfluid phase-space density, $\Ds = \ns \lambda^2$, deduced from the measured sound speeds. Here $\mathcal{D}=n\lambda^2$ is the total phase space density and $\Dc$ is its critical value. 
The solid line, showing the universal jump of $\Ds$ from $0$ to $4$ at $\mathcal{D}=\Dc$, is the infinite-system theoretical prediction without any free parameters. The dashed line corresponds to a $100\%$ superfluid ($\Ds = \mathcal{D}$). All error bars show standard statistical errors.
 }
\label{fig:3}
\end{figure*} 

In Fig.~\ref{fig:2}c we plot the fitted $\fsum$ for a range of temperatures, showing that it always satisfies the $f$ sum rule. The bottom panel shows $f_2$, the second-sound contribution to the constant $\fsum$, which vanishes with increasing $T$ and we use it to experimentally identify $\Tc$. In absolute terms, including our systematic uncertainties in $n$ and $T$ (see Methods), we get $\Tc = 42(4)$~nK for $n = 3.0(5)~\mu$m$^{-2}$, which is compatible with the BKT prediction~\citep{Prokofev:2001} $\Tc = 2\pi n\hbar^2/[m\kB \ln{(380/\g)}]= 37(6)$~nK. For our system parameters, finite-size effects might indeed shift $\Tc$ up by $\sim 10\%$ (see Refs.~\citep{Pilati:2008,Foster:2010, Hung:2011, Gawryluk:2019}), but within our systematic errors this shift is not conclusive. 

In the BKT theory for an infinite system, at $\Tc$ the superfluid density $\ns$ exhibits a jump from $0$ to $4/\lambda^2$, where $\lambda$ is the thermal wavelength~\citep{Nelson:1977}. This in turn implies that, in contrast to the 3D case, $c_2$ and $f_2$ are discontinuous at $\Tc$~\citep{Ozawa:2014}. 
In Fig.~\ref{fig:2}c (bottom) the value of the theoretical jump in $f_2$ is indicated by the dashed line~\citep{Ota:2018b}. In a finite-size system the transition is necessarily rounded-off into a crossover \citep{Pilati:2008,Foster:2010, Gawryluk:2019} (see also~\citep{Bishop:1978,Wu:2020} for nonzero-$\omega$ effects), but nevertheless the variation of $f_2$ near $\Tc$ is quite sharp.

In Fig.~\ref{fig:3}a we show our results for the speeds of first (red) and second (blue) sound, together with the infinite-system predictions ~\citep{Ozawa:2014,Ota:2018b,Prokofev:2002} without any free parameters. The theoretical normalised speeds $c_{1,2}/\cB$ are functions of just $T/\Tc$ and $\g$, based on the thermodynamic results of Ref.~\citep{Prokofev:2002}. In our case, the superfluid-density jump at $\Tc$ corresponds to a discontinuity in $c_2$ of $ \approx0.45\,\cB$; the theory also predicts a smaller ($ \approx 0.1\,\cB$) discontinuity in $c_1$, which is comparable to our experimental errors. We note that the calculations in Ref.~\citep{Prokofev:2002} were done assuming $\g \ll 1$, but they are expected to be applicable to our $\g$~\citep{Ha:2013}, as long as $\ns$ is notably below the total $n$~\citep{Ota:2018b}. For $0.5 \lesssim T/\Tc \lesssim 0.75$ the predictions are not reliable for $c_1$, which is very sensitive to the exact value of the vanishingly small normal-fluid density, while the predictions for $c_2$, which is simply $\approx \cB$, are more robust~\citep{Ota:2018b}.

Finally, following Ref.~~\citep{Ozawa:2014}, we deduce the superfluid density from our measured sound speeds and the universal, scale-invariant 2D equation of state~\citep{Prokofev:2001,Prokofev:2002, Yefsah:2011,Hung:2011,Ha:2013} (see Methods). 
While in the strongly interacting 3D Fermi superfluids $\ns$ can be deduced, to a good approximation, from $c_2$ alone~\citep{Sidorenkov:2013}, in our case the deduced $\ns$ is, in the most interesting region just below $\Tc$, very sensitive to the values of both $c_1$ and $c_2$.
In Fig.~\ref{fig:3}b, we plot the superfluid phase-space density $\Ds = \ns \lambda^2$ versus $\mathcal{D}/\Dc$, where $\mathcal{D} = n\lambda^2$ is the total phase-space density and $\Dc$ is its critical value. The solid line is the theoretical prediction~\citep{Prokofev:2002} without any free parameters, which shows the universal, $\g$-independent jump of $\Ds$ from $0$ to $4$ at the critical point.
The dashed line corresponds to a pure superfluid ($\Ds = \mathcal{D}$).

Our experiments establish the applicability of the celebrated two-fluid model to the unconventional BKT superfluids, and provide the first measurement of the superfluid density in an atomic 2D gas, showing agreement with its expected universal jump at the critical point. 
Our measurements also extend into the low-temperature regime where reliable predictions are not available, and thus provide guidance for further theoretical work. An experimental challenge for the future is to explore even lower temperatures, where hybridization of the first and second sound is expected~\citep{Lee:1959,Stringari:2017}.
More generally, the establishment of the measurements of the superfluid density in 2D quantum gases provides a new invaluable diagnostic tool for many future studies, including explorations of non-equilibrium phenomena and the effects of disorder on superfluidity.


We thank Jay Man for experimental assistance, and Robert P. Smith, Jean Dalibard, Martin Zwierlein, Richard J. Fletcher, Timon A. Hilker, and Sylvain Nascimbene for fruitful discussions. This work was supported by EPSRC [Grants No.~EP/N011759/1 and No.~EP/P009565/1], ERC (QBox) and QuantERA (NAQUAS, EPSRC Grant No.~EP/R043396/1). J.~S. acknowledges support from Churchill College (Cambridge). Z.~H. acknowledges support from the Royal Society Wolfson Fellowship.


%

\setcounter{figure}{0} 
\setcounter{equation}{0} 

\renewcommand\theequation{S\arabic{equation}} 
\renewcommand\thefigure{S\arabic{figure}} 

\section*{Methods}

{\bf Optical confinement of the 2D gas.}
The 1D optical lattice and the rectangular hollow beam are blue-detuned from the atomic resonance, and create repulsive potentials for the atoms. Both are shaped using digital micromirror devices (DMDs). The hollow beam providing the in-plane confinement has a wavelength of $760$~nm and is created by direct imaging of a DMD pattern. The vertical 1D lattice is made of $532$-nm light and is created by Fourier imaging of a DMD pattern, which allows dynamical tuning of the lattice period $\Delta z$. Specifically, using a DMD we create two horizontal light strips, each of width corresponding to $50$ micromirror pixels, separated vertically by $\Delta Z$, so their interference in the Fourier plane creates a lattice with $\Delta z \propto 1/\Delta Z$. We additionally impose a $\pi$ phase shift between the two interfering beams, which places the central node of the symmetric interference pattern at $z=0$ independently of the varying $\Delta z$. To dynamically change $\Delta z$ we shift the DMD pattern pixel by pixel (moving two light strips symmetrically in opposite directions) in $25$-ms steps. 
We start with a large $\Delta z = 18.5~\mu$m to load a pre-cooled 3D gas~\citep{Campbell:2010} into a single lattice node, and then reduce it over $1.5$~s to $3.3~\mu$m in order to compress the gas into the 2D geometry.
In the final 2D configuration, the lattice depth around the central node is $U_z \simeq 3.0~\mu$K, giving the trap frequency $\omega_z/(2\pi)=(\Delta z)^{-1} \sqrt{U_z/(2m)}=5.5$~kHz.

\vspace{2mm}

{\bf Calibration of the experimental parameters.}
Our absorption-imaging system, used to measure the cloud density $n$, is calibrated with a systematic uncertainty of $15\%$ using measurements of the critical temperature for Bose--Einstein condensation in a 3D harmonic trap~\citep{Campbell:2010}; this calibration also agrees with an independent calibration based on the rates of the density-dependent three-body decay~\citep{Zaccanti:2009}. We assess the absolute gas temperature with a systematic uncertainty of $10\%$ using measurements of the scale-invariant 2D equation of state (EoS)~\citep{Prokofev:2002,Yefsah:2011,Ha:2013}, as in~\citep{Ville:2018}; we have made EoS measurements for several trap depths $U_0$ and also different trap dimensions $\Lx$ and $\Ly$, which show linear dependence of $T$ on $U_0$.
The wavevector $q=\pi/\Ly$ is determined using {\it in situ} absorption images (such as shown in Fig.~\ref{fig:1}b), with a systematic $5\%$ error due to the fact that the cloud edges are not infinitely sharp; the half-wavelength of the density oscillations closely corresponds to the length of the region in which the density is above $90\%$ of its value in the bulk. The driving force $F_0$ is calibrated with an error of $5\%$ by applying a constant force on a cloud released from the trap and measuring the resulting centre-of-mass acceleration.

\vspace{2mm}

{\bf Response function, \texorpdfstring{$f$}{} sum rule, and \texorpdfstring{$S(\omega)$}{}.}
The density response function is defined in Fourier space as $\xn (q, \omega) = \delta n(q,\omega)/\delta U(q,\omega)$, where $\delta U(q,\omega)$ is the driving potential. Our monochromatic and spatially uniform driving force corresponds to a potential $-F_0 y \sin(\omega t)$ for $-\Ly/2 \leq y \leq \Ly/2$, and Fourier decomposing this gives $\delta U (q = \pi/\Ly, \omega) = - 4 F_0 \Ly/\pi^2$. Following our definition of $A(\omega)$ gives $\Ima[\xn (q = \pi/\Ly,\omega)] = - \pi^2 nq^2 A(\omega) / (8F_0)$. Inserting this into the conventional form of the $f$ sum rule~\citep{Pitaevskii:2016},
\begin{equation}
\int_{-\infty}^{\infty} {\rm d}\omega \, \omega \, \Ima[\xn (q,\omega)] = - \frac{\pi nq^2}{m} \, ,
\end{equation}
gives the dimensionless sum rule in Eq.~(\ref{eq:f}),
which is insensitive to uncertainties and variations in $n$ and $q$. Our dimensionless results in Fig.~\ref{fig:2}c (and also Fig.~\ref{fig:3}) are also not affected by changes in $F_0$; for these different measurements we have varied $F_0$ by a factor of $3$, and also $q$ by about $50\%$ by choosing various box sizes.
The dynamical structure factor is (for $\kB T \gg \hbar \omega$, which is always satisfied in our experiments) given by $S(q, \omega) = - \kB T \, \Ima[\xn (q,\omega)] / (\pi n \omega)$, which is equivalent to the form given in the main text in terms of $A(\omega)$.

\vspace{2mm}

{\bf Fits of the response spectra and extraction of \texorpdfstring{$\ns$}{}.}
In the two-fluid model, the two sound speeds, $c_1$ and $c_2$, are solutions of the quartic equation for $c$:
\begin{equation}
   c^4-(c_{10}^2+c_{20}^2)c^2+c_{10}^2c_{20}^2/\gamma = 0 \, ,
\end{equation}
where $c_{10}^2=1/(m n\kS)$ and $c_{20}^2=Ts^2\ns/[m c_V(n-n_{\text{s}})]$; here $c_V$ is the specific heat per particle at constant volume, $s$ is the entropy per particle, and $\gamma = \kT/\kS$ is the ratio of the isothermal and isentropic compressibilites.
Due to the scale-invariance in 2D, for a given $\g$ all these thermodynamic quantities depend only on $T/\Tc$ through the dimensionless phase-space density and pressure~\citep{Prokofev:2002,Ozawa:2014}. Conversely, experimentally measuring $c_1$ and $c_2$ (for $T<\Tc$) gives $c_{20}$ and hence $\ns$. Ignoring dissipation, the density response function is~\citep{Hu:2010}
\begin{equation}
    \xn(q, \omega) =\frac{nq^2}{m}\left(\frac{Z_1}{\omega^2-c_1^2q^2}+\frac{Z_2}{\omega^2-c_2^2q^2}\right) \, ,
    \label{response}
\end{equation}
with the two poles giving $c_{1,2}$, and $Z_1+Z_2 = 1$ to satisfy the $f$ sum rule. 
Including linear damping~\citep{Hohenberg:1965}, we fit the experimental spectra with $A(\omega) = A_1(\omega)+A_2(\omega)$, where
\begin{equation} 
  A_{1,2}(\omega) = \frac{x_{1,2} \omega_{1,2}^2\Gamma_{1,2}\, \omega}{(\omega^2-\omega_{1,2}^2)^2 +(\omega\, \Gamma_{1,2})^2} \, . 
\end{equation}
Here the amplitudes $x_{1,2}$, resonance frequencies $\omega_{1,2}$, and damping rates $\Gamma_{1,2}$, with $\omega_1 > \omega_2$, are fit parameters, and the sound diffusivities are then given by $\Gamma_{1,2}/q^2$. For consistency, we first apply the same fit to the data taken at all temperatures, and find that it always captures the data well and gives $\fsum \approx 1$ (see Fig.~\ref{fig:2}c). The first sound is always underdamped and the $A_1$ term gives $f_1$, its contribution to $\fsum$. For the spectra identified as being below $\Tc$ (as in the top panel of Fig.~\ref{fig:2}a), the fit gives that the second sound is also underdamped, and its contribution to $S(\omega)$ peaks at a nonzero $\omega$. In this case $A_2$ gives the nonzero $f_2$ contribution to $\fsum$. For the data identified as being above $\Tc$ (as in the bottom panel of Fig.~\ref{fig:2}a), the second term in the fit function gives that this mode is overdamped, and its contribution to $S(\omega)$ peaks at $\omega=0$. This shows in an unbiased way that the second sound is replaced by the diffusive mode. In this case the $A_2$ term gives $f_{\rm diff}$, the diffusive-mode contribution to $\fsum$, with $f_1 + f_{\rm diff} \approx 1$, while $f_2=0$.  
To estimate the thermal diffusivity, following~\citep{Hohenberg:1965,Hu:2018} we refit the data for $T>\Tc$ with $A(\omega) = A_1(\omega)+A_{\text{T}}(\omega)$, where  
\begin{equation} 
       A_{\text{T}}(\omega)=\frac{x_{\text{T}}\, \Gamma_{\text{T}}\,\omega}{\omega^2+\Gamma_{\text{T}}^2} \,
\end{equation}
corresponds to $S(\omega)$ contribution that is a Lorentzian centered at $\omega = 0$, and gives $D_{\text{T}}=\Gamma_{\text{T}}/q^2$.

\end{document}